# Majority-Vote Cellular Automata, Ising Dynamics, and P-Completeness

Cristopher Moore

Santa Fe Institute
1399 Hyde Park Road, Santa Fe NM 87501 USA
moore@santafe.edu

**Abstract.** We study cellular automata where the state at each site is decided by a majority vote of the sites in its neighborhood. These are equivalent, for a restricted set of initial conditions, to non-zero probability transitions in single spin-flip dynamics of the Ising model at zero temperature.

We show that in three or more dimensions these systems can simulate Boolean circuits of AND and OR gates, and are therefore **P**-*complete*. That is, predicting their state $t$ time-steps in the future is at least as hard as any other problem that takes polynomial time on a serial computer. Therefore, unless a widely believed conjecture in computer science is false, it is impossible even with parallel computation to predict majority-vote cellular automata, or zero-temperature single spin-flip Ising dynamics, qualitatively faster than by explicit simulation.

## 1 Introduction

The "complexity" of a physical system can be defined as the amount of time, memory, or other computational resources needed to predict it. How these resource requirements vary with the system size, or the number of time-steps we wish to predict, becomes increasingly important as we study larger and larger systems for longer and longer times. For instance, a computation time of $\mathcal{O}(\log t)$ or $\mathcal{O}(\log^2 t)$ would be greatly preferable to $\mathcal{O}(t)$ for large $t$.

For example, consider a cellular automaton (CA), a dynamical system with a discrete set of local states and a local update rule where each site's evolution depends on its state and those of its neighbors. We can predict the state of a given site $t$ time-steps in the future by explicitly simulating the CA and filling in a light-cone of depth $t$ ending in that site. This takes time $\mathcal{O}(t^{d+1})$ on a serial computer (proportional to the volume of the light-cone in $d$ dimensions) or just $\mathcal{O}(t)$ if done in parallel.

However, for classes of CAs that obey certain algebraic identities, we can do much better than this [6, 7]. These "quasi-linear" systems can be predicted on a parallel computer in time $\mathcal{O}(\log t)$ or $\mathcal{O}(\log^2 t)$, much faster than by explicit simulation. This places them in the complexity class **NC**, the class of problems that can be solved for inputs of size $n$ in time $\mathcal{O}(\log^k n)$ for some $k$, on a parallel computer with a polynomial number $\mathcal{O}(n^c)$ of processors. In this case, the input consists of $n = \mathcal{O}(t^d)$ initial sites at the top of the light-cone.

**NC** is a subset of **P**, the class of problems solvable in polynomial time on a serial computer [3, 9]. It is believed but not known that there are *inherently sequential* problems, in **P** but not in **NC**, where a substantial amount of work has to be done in step-by-step order, so that even a large number of parallel processors cannot solve them in polylogarithmic time. In our CA example, this would correspond to a rule for which no fast parallel algorithm exists, so that the fastest method of prediction is to simulate them explicitly.

**P**-*complete* problems, to which any other problem in **P** can be reduced, are the most likely to be inherently sequential. If any **P**-complete problem has a fast parallel algorithm, then every problem in **P** does, in which case **P** = **NC**. (Similarly, if an **NP**-complete problem such as the Traveling Salesman Problem has a deterministic polynomial-time algorithm, then **NP** = **P**.)

One such problem is CIRCUIT VALUE. Given a Boolean circuit, i.e. a directed graph whose nodes are AND, OR and NOT gates, and given the truth values of its inputs, is the output true or false? This is **P**-complete since any deterministic Turing machine computation of length $t$ can be converted to a Boolean circuit of depth $\mathcal{O}(t)$; thus polynomial-time computations are equivalent to polynomial size (and depth) circuits. Since the truth values at each level of the circuit affect those on the next level in arbitrary ways, it is hard to imagine how one could calculate the output without going sequentially through the circuit level by level; an **NC** algorithm would have to somehow evaluate many levels at once, or provide a method of skipping over most of them. (In fact, **P** = **NC** if and only if circuits of polynomial depth always have much shallower equivalent circuits.)

The MONOTONE CIRCUIT VALUE problem, where AND and OR gates are allowed but NOT gates are not, is still **P**-complete for the following reason: using De Morgan's laws $\overline{a \wedge b} = \overline{a} \vee \overline{b}$ and $\overline{a \vee b} = \overline{a} \wedge \overline{b}$, we can shift negations back through the gates until they only affect the inputs themselves. Thus any CIRCUIT VALUE problem can be converted to a MONOTONE CIRCUIT VALUE problem with some of the inputs negated. This kind of conversion, from an instance of one problem to an instance of another, is called a *reduction*.

We will show that two cellular automaton rules based on voting among neighboring sites are just as hard to predict as MONOTONE CIRCUIT VALUE is to solve, by showing how to build arbitrary monotone Boolean circuits into the CAs' initial conditions. Thus these CAs are **P**-complete, and unless **P** = **NC** there is no way to predict them much faster than explicit simulation.

We will then relate these CAs to single spin-flip dynamics of the zero-temperature Ising model, and show that the question of whether a given spin will be "up" with non-zero probability is also **P**-complete. Our results hold for $d \geq 3$.

To our knowledge, the first use of **P**-completeness in statistical physics is by Machta and Greenlaw [5], who show that several non-deterministic systems, including diffusion-limited aggregation and Ising dynamics at non-zero temperatures, are **P**-complete when a series of random bits directing the system's non-deterministic choices (how a particle diffuses, or which spin will try to flip) is included in the input. Greenlaw et al. [3] have pointed out that predicting a CA's evolution is **P**-complete in general, since CA rules exist (e.g. [4]) which can simulate universal Turing machines.

## 2 Majority-Vote Cellular Automata

Let us consider the following CA rule, with two local states $\{0, 1\}$:

$$s'_i = \begin{cases} 1 \text{ if } s_i + \sum_j s_j > d \\ 0 \text{ otherwise} \end{cases}$$

where in $d$ dimensions the sum is over the $2d$ nearest neighbors of $s_i$ in the cubic lattice. In other words, $s_i$'s next state is decided by a majority vote of the $2d+1$ sites in its immediate neighborhood, including itself (this is sometimes called the Von Neumann neighborhood). We will call this the *majority-vote CA*.

To be precise about the problem we're posing, define MAJORITY-VOTE CA PREDICTION as the following: given initial conditions in a volume with $V$ sites, will a certain site be occupied after $t$ time-steps where $t \leq V$? Then we wish to show the following:

**Theorem 1.** MAJORITY-VOTE CA PREDICTION *is* **P**-*complete for* $d \geq 3$.

*Proof.* We can clearly solve this problem in **P** by explicit simulation, since $t$ is bounded by the length of the input $V$; in fact, any polynomial bound $t < \mathcal{O}(V^k)$ would ensure this. To show that it is **P**-complete we will reduce MONOTONE CIRCUIT VALUE to it.

In figure 1 we show how to build AND and OR gates in three dimensions; there are probably many such constructions. We connect them with 'wires' with a U-shaped cross-section. The reader can easily check that the occupied sites of the wire (those with $s = 1$) have four or five occupied sites in their neighborhood, including themselves; the sites in the trough of the wire have three occupied sites in their neighborhood, one short of a majority. Therefore, the wire is fixed under the CA rule.

If one site in the trough becomes occupied, the site next to it will have four occupied neighbors, and will become occupied at the next time-step; this will 'turn on' the site next to it, and so on. Thus the wire will propagate this signal (and, as we show below, it will never turn off after turning on). We will use this to represent truth in the network.

The AND and OR gates work in a simple way. Each has three wires meeting at a central site. For the AND gate, this central site has two occupied neighbors, so if any two of the input wires turn on, it gains a majority and sends a signal out along the third wire. In the OR gate the central site has three occupied neighbors, so if any one of the three inputs turns on, it gains a majority and turns on the other two. In addition, every occupied site in these gates has at least four occupied neighbors, so the gates are fixed under the rule.

OR gates can also be used to make a wire branch, so that the output of a gate can 'fan out' and be input into any number of others.

Note that timing is not an issue here: if a single input to an AND gate turns on, for instance, it waits there patiently to see if it will be joined by another.

Finally, to connect these gates we need to be able to bend the wires to match the gates' inputs. Figure 1 shows two "elbows" which can be combined to re-orient a wire in any of its 24 possible directions and orientations; straight sections can then extend it from place to place as desired.

These constructions allow us to convert any monotone Boolean circuit to a set of initial conditions in some volume $V$, such that after at most $V$ time-steps (in fact, after a number of time-steps bounded by the total number of sites in the wires' troughs) a site representing the output will become occupied.

So MAJORITY-VOTE CA PREDICTION in three dimensions is **P**-complete, since MONOTONE CIRCUIT VALUE can be reduced to it; if the majority-vote CA can be predicted in $\mathcal{O}(\log^k t)$ parallel time, then we can use it to solve monotone circuits just as quickly, and **P** and **NC** are equal.

We next show how to simulate the $d = 3$ majority vote CA in higher dimensions. We claim that for every $b$ there is a $b$-dimensional configuration on the infinite lattice $\mathbb{Z}^b$ such that the site at the origin $(0, \ldots, 0)$ has half its neighbors occupied (not including itself) and half unoccupied, and all other sites are fixed under the CA rule.

We will show this by induction as in figure 2. For $b = 0$, we have a single site with no neighbors, half of which are occupied and half of which are unoccupied!

Then given such a configuration $f_{b-1}$ on $\mathbb{Z}^{b-1}$ we can make one $f_b$ on $\mathbb{Z}^b$ by writing

$$f_b(x_1, x_2, \ldots, x_b) = \begin{cases} 1 & x_b < 0 \\ f_{b-1}(x_1, x_2, \ldots, x_{b-1}) & x_b = 0 \\ 0 & x_b > 0 \end{cases}$$

The reader can easily show that sites with $x_b \neq 0$ have at most one neighbor unlike themselves, so they are fixed under the CA rule. A site with $x_b = 0$ has one neighbor occupied and one unoccupied (with $x_b = \pm 1$) so its majority status is determined by its $2(b-1)$ neighbors with $x_b = 0$. But these are described by $f_{b-1}$ unless $x_{b-1} = 0$, and so on. Finally, the site at $(0, \ldots, 0)$ has $b$ occupied neighbors and $b$ unoccupied, being those whose single non-zero $x_i$ is $-1$ or $+1$ respectively.

Then to simulate the evolution of some three-dimensional configuration $g$ in $d \geq 3$ dimensions, let $b = d - 3$ and define

$$g'(x_1, x_2, x_3, y_1, y_2, \ldots, y_b) = \begin{cases} g(x_1, x_2, x_3) & \text{if } y_i = 0 \text{ for all } i \\ f_b(y_1, \ldots, y_b) & \text{otherwise} \end{cases}$$

Then the evolution of $g'$ on a $d$-dimensional lattice will contain, at $y_i = 0$ for all $i$, the evolution of $g$ on the three-dimensional lattice. So the $d = 3$ case can be simulated by any $d \geq 3$, and MAJORITY-VOTE CA PREDICTION is **P**-complete for $d \geq 3$. □

We can do a similar construction for a slightly different CA rule:

$$s_i' = \begin{cases} 1 \text{ if } \sum_j s_j \geq d \\ 0 \text{ otherwise} \end{cases}$$

Here $s_i$ will be occupied on the next time step if $d$ or more of its $2d$ nearest neighbors, not including itself, are occupied. We will call this the *half-or-more CA*.

**Theorem 2.** HALF-OR-MORE CA PREDICTION *is* **P**-*complete for* $d \geq 3$.

*Proof.* We give a construction in figure 3. Our wires have a V-shaped intersection, with occupied sites having three or four occupied neighbors and sites in the trough having two. Three such wires can come together in a pleasant intersection. By adding $2 \times 2 \times 2$ blocks, which are fixed under the rule, we can give the center site one or two occupied neighbors, creating an AND or OR gate respectively. Because of the diagonal symmetry of the wire, only one kind of elbow is required to bend it into any desired direction and orientation. Higher dimensions can simulate the $d = 3$ case as in theorem 1. □

## 3 Ising Dynamics

The *Ising model* is an idealization of magnetic materials in which each site in a lattice has a spin $s_i = \pm 1$. The total energy is then

$$E = -J \sum_{i,j} s_i s_j$$

summed over all pairs of nearest neighbors. In a *ferromagnetic* (resp. *anti-ferromagnetic*) material, $J > 0$ ($J < 0$) and $H$ is minimized when adjacent sites have the same (opposite) spin. On any two-colorable lattice, we can change $J$ to $-J$ by choosing a sub-lattice of sites and flipping all their spins; thus for cubic lattices, ferromagnetic and anti-ferromagnetic systems are equivalent.

In single spin-flip dynamics, at each time-step we choose a random site in the lattice and seek to minimize its contribution to the total energy from its interaction with its nearest neighbors $-J s_i \sum_j s_j$. If the system is exposed to a heat bath at temperature $T$, states with energy $E$ are chosen with probability proportional to $e^{-E/kT}$ where $k$ is Boltzmann's constant; at $T = 0$ lower energy states are always preferred. Thus if a majority of $s_i$'s neighbors are up (down), we will change it to up (down); if its neighbors are half up and half down, we will flip it with probability $1/2$.

We can then formulate a prediction problem, called ZERO-TEMPERATURE SINGLE SPIN-FLIP ISING DYNAMICS: given an initial set of spins, is there a non-zero probability that a particular spin $s_i$ will be $+1$ after $t$ time-steps? That is, does there exist a sequence of sites (and a sequence of choices of whether or not to flip sites whose neighbors sum to zero) that leads to $s_i = +1$?

We will show this problem is **P**-complete by reducing the special case of HALF-OR-MORE CA PREDICTION we constructed above to it. We use the following lemma:

**Lemma 3.** *For the choice of initial conditions constructed in theorem 2, a site will eventually become occupied under the half-or-more CA rule if and only if there is a non-zero probability of that site flipping up under zero-temperature single spin-flip Ising dynamics (where occupied and unoccupied sites correspond to up and down spins respectively).*

*Proof.* Let $S_0$ be the set of sites that are occupied in the initial conditions. Then for all $k \geq 0$, let $S_{k+1}$ be the set of sites with $d$ or more neighbors in the union $\cup_{i \leq k} S_k$ of the previous sets. Call the union of all of these $\overline{S} = \cup_{k \geq 0} S_k$.

It is clear that $\overline{S}$ is precisely the set of sites with a non-zero probability of flipping up. Any such site must have $d$ neighbors who have a non-zero probability of flipping up before it does, and these must have neighbors before them, back to the initial sites $S_0$. Conversely, any site $s$ in $\overline{S}$ can be flipped by flipping all the sites in $S_1 - S_0$, then all of $S_2 - (S_0 \cup S_1)$, and so on until we get to it, flipping each site once (always choosing to flip up in case of a tie).

Now for the initial conditions we constructed in theorem 2, every site in $S_0$ has $d$ or more neighbors in $S_0$; thus sites in $S_0$ will keep each other occupied. Call such a set *self-sustaining*. Inductively, if after $k$ time-steps the set of occupied sites is $S_k$, and if $S_k$ is self-sustaining, then $S_k \subset S_{k+1}$ by definition, $S_{k+1}$ is self-sustaining also, and all of $S_{k+1}$ will be occupied on the next time-step.

Thus $\overline{S}$ is also precisely the set of sites that the half-or-more CA will turn on when starting with self-sustaining initial conditions, and the lemma is proved. □

Then we have

**Theorem 4.** Zero-Temperature Single Spin-Flip Ising Dynamics *is* **P**-*complete.*

*Proof.* We have reduced Monotone Circuit Value to Half-or-More CA Prediction, and reduced it in turn (in the special case of self-sustaining initial conditions) to Zero-Temperature Single Spin-Flip Ising Dynamics. The output of the circuit is true if and only if a particular site has a non-zero probability of flipping up. □

## 4 Discussion and Conclusion

We have shown that several systems are **P**-complete to predict in three or more dimensions: majority-vote cellular automata, half-or-more cellular automata, and single spin-flip Ising dynamics at zero temperature. Unless $\mathbf{P} = \mathbf{NC}$, then, these systems are inherently sequential, and no speedup to polylogarithmic time is possible.

(Even if $\mathbf{NC} < \mathbf{P}$, a power-law speedup might still be possible; for instance, Moriarty and Machta give an algorithm for diffusion-limited aggregation with an expected power-law speedup [8]. We conjecture, however, that for voting CAs and Ising dynamics in $d \geq 3$, no power-law speedup beyond $\mathcal{O}(t)$ is possible.)

In one dimension, all these problems are easy. In the nearest-neighbor majority-vote CA, any domain of 0s or 1s of length 2 or more is stable, and the boundaries between them are fixed; checkerboards of alternating 0s and 1s collapse at the speed of light. In Ising dynamics (or the half-or-more CA), every site can flip up (become occupied) unless the initial conditions are all down (unoccupied).

What about two dimensions? Monotone Boolean circuits that are also *planar* [1, 2] can be evaluated in **NC**. It seems possible, then, that there is a fast algorithm to predict these systems for $d = 2$. In the case of Ising dynamics, this would imply that a system's complexity class is not necessarily related to its thermodynamic behavior: a system with a finite thickness in a third dimension can embed non-planar graphs and thus render the problem **P**-complete, while such a system is in the same thermodynamic universality class as the $d = 2$ Ising model.

**Acknowledgements.** I am indebted to Jonathan Machta, Mats Nordahl, Richard Beigel, Raymond Greenlaw, Manor Askenazi and Joshua Berman for stimulating conversations, to Elizabeth Hunke and Spootie the Cat for companionship, and to the Santa Fe Brewing Company for inspiration.

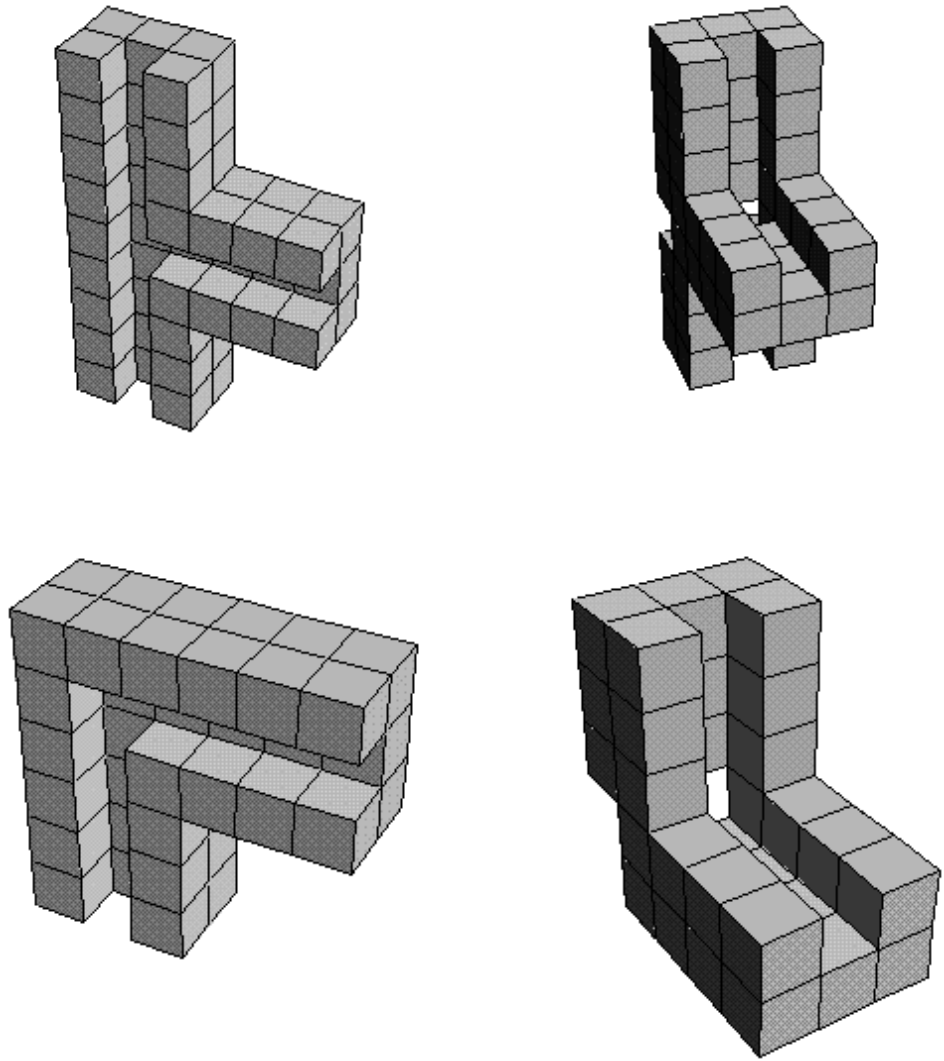

**Fig. 1.** Clockwise from upper left, an AND gate, OR gate, and two types of elbows for bending wires in the majority-vote CA. The trough connecting the vertical wires in the OR gate goes all the way through, and the 'notch' extends through all three sites in the back.

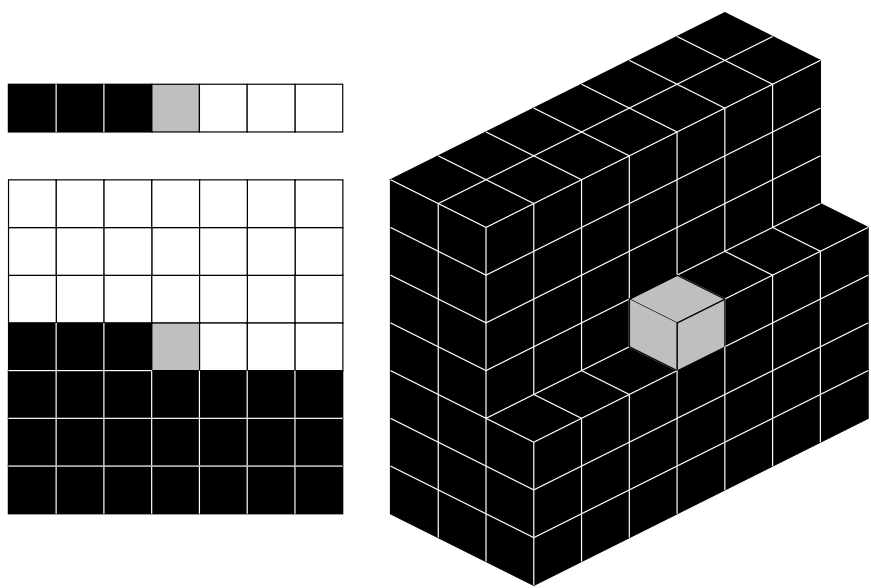

**Fig. 2.** One, two, and three-dimensional configurations where the site at the origin (grey) has half its neighbors occupied and half unoccupied, and all other sites have a majority like themselves and so are fixed under the CA rule.

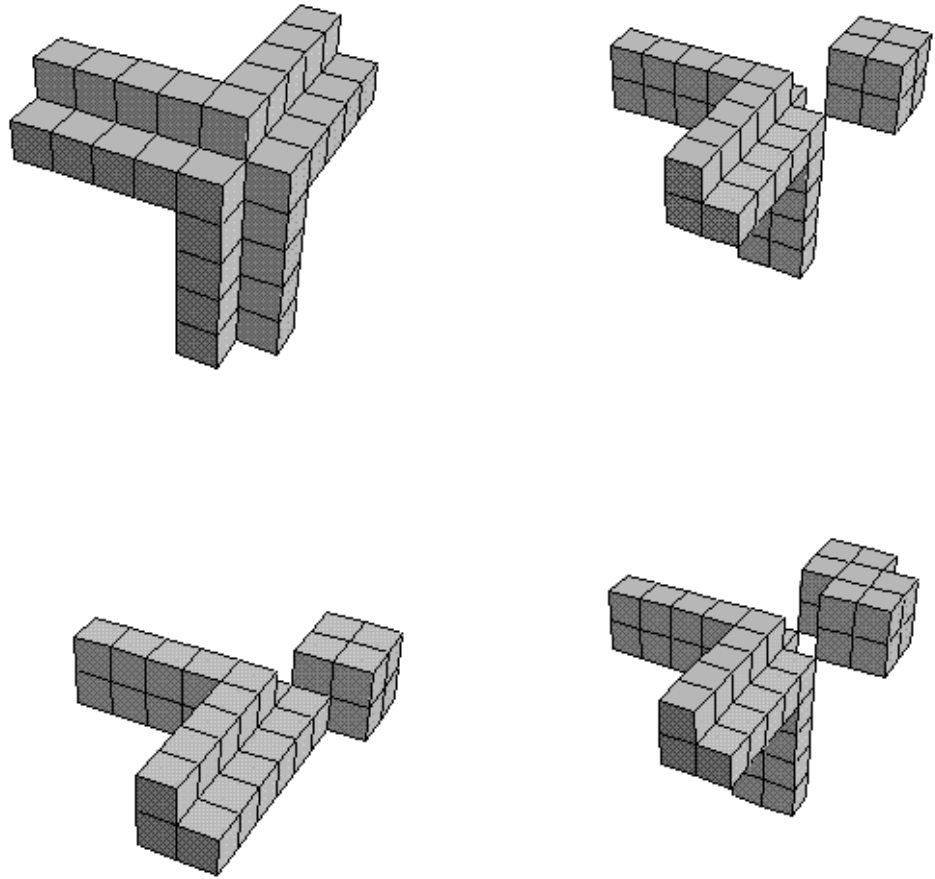

**Fig. 3.** Clockwise from upper left, the basic junction of three wires in the half-or-more CA; AND and OR gates formed by placing one or two $2 \times 2 \times 2$ blocks adjacent to the center site; and an elbow.